\documentclass[journal]{IEEEtran}
\usepackage[utf8]{inputenc}
\usepackage{amsmath}
\usepackage{amssymb}
\usepackage{graphicx}
\usepackage{float}
\usepackage{subeqnarray}
\usepackage{cases}
\usepackage{color}
\allowdisplaybreaks
\IEEEoverridecommandlockouts
\usepackage[square, comma, sort&compress, numbers]{natbib}
\usepackage{multirow}
\usepackage{stfloats}
\usepackage{fixltx2e}
\usepackage[font={small}]{caption}
\usepackage{epsfig}
\usepackage{epstopdf}
\usepackage{algorithm}
\usepackage{algpseudocode}
\usepackage{amsmath}

\newtheorem{Theorem}{Theorem}

\newtheorem{Corollary}{Corollary}

\begin{document}
\title{Downlink Asynchronous Non-Orthogonal
Multiple Access with Quantizer Optimization}

\author{\IEEEauthorblockN{Xun Zou, \textit{Student Member, IEEE}, Mehdi~Ganji, \textit{Student Member, IEEE}, and Hamid~Jafarkhani, \textit{Fellow, IEEE}}\\
	\thanks{The authors are with the Center for Pervasive Communications and Computing, Department of Electrical Engineering and Computer Science, University of California, Irvine, CA, 92697 USA (email: \{xzou4, mganji, hamidj\}@uci.edu). This work was supported in part by the NSF Award CCF-1526780.}
}
\maketitle
\begin{abstract}
    In this letter, we study a two-user downlink asynchronous non-orthogonal multiple access (ANOMA) with limited feedback. We employ the max-min criterion for the power allocation and derive the closed-form expressions for the upper and lower bounds of the max-min rate. It is demonstrated that ANOMA can achieve the same or even higher average max-min rate with a lower feedback rate compared with NOMA. Moreover, we propose a quantizer optimization algorithm which applies to both NOMA and ANOMA. Simulation results show that the optimized quantizer significantly improves the average max-min rate compared with the conventional uniform quantizer, especially in the scenario with a low feedback rate.
\end{abstract}

\section{Introduction}
Non-orthogonal multiple access (NOMA) has been regarded as one of the key technologies to meet the challenges of the next generation wireless communications. The key feature of NOMA is that different users' signals can share the same time and frequency resources. The NOMA consists of power-domain NOMA, code-domain NOMA, and the joint design of both~\cite{zou2019TCNOMA}. In power-domain NOMA, the superposition coding and the successive interference cancellation (SIC) are utilized for multi-user transmission and detection, respectively.

Recently, a novel scheme called asynchronous NOMA (ANOMA) has been proposed to further improve the performance of NOMA, for example, in the uplink system~\cite{zou2019analysis} and the cooperative network~\cite{zou2019cooperative}. In ANOMA, the intentional introduction of timing mismatch at the transmitter and the oversampling technique at the receiver result in the sampling diversity~\cite{zou2019analysis,zou2019cooperative,ganji2019improving}. It has been demonstrated that ANOMA outperforms NOMA in terms of the throughput performance, the power consumption, etc. Moreover, the time asynchrony has also been exploited in multi-user transmit beamforming~\cite{mehdi2020exploiting}.

The channel state information (CSI) plays a critical role in optimizing the system performance. 
At the transmitter side, the CSI is employed to conduct the adaptive power/rate allocation and generate the beamforming vectors in multiple-input multiple-output (MIMO) systems. In time-division duplexing (TDD) systems, the channel reciprocity is exploited by the base station (BS) to utilize the CSI estimated via the uplink training. In frequency-division duplexing (FDD) systems, a prevailing technique is using the limited feedback from users. The NOMA with limited feedback has been studied in the existing literature, for example, the one-bit feedback scheme in the massive MIMO NOMA systems~\cite{ding2016design} and the multi-user single antenna systems~\cite{xu2016outage}, and the scalar quantizer design in downlink power-domain NOMA~\cite{liu2017downlink}. To the best of our knowledge, the analysis of limited feedback schemes in ANOMA systems and the optimal quantizer design for NOMA/ANOMA are still absent. In fact, the limited feedback design in NOMA/ANOMA systems is more challenging compared with that in the orthogonal multiple access (OMA) scenario, e.g., in~\cite{karamad2012quantization}. It is because the CSI of each user not only affects its own but also other users' performance due to the inter-user interference (IUI) ingrained in the non-orthogonal transmission. In more details, the CSI is used for both allocating powers and determining the SIC order, which further complicates the rate expressions and the system design.

In this letter, we consider a downlink ANOMA system with limited feedback. We employ the max-min criterion for the power allocation and the scalar quantizer for channel quantization, respectively. We derive the closed-form expressions for the upper and lower bounds of the max-min rate. It is manifested that ANOMA can achieve the same or even higher average max-min rate with a lower feedback rate compared with NOMA. Moreover, we propose a quantizer optimization method which applies to both NOMA and ANOMA systems. A gradient descent algorithm is designed to optimize the quantization levels. Simulation results show that a higher average max-min rate is achieved by using the optimized quantizer compared with the conventional uniform quantizer in~\cite{liu2017downlink}, especially for the low-rate feedback scenario.

\section{Preliminaries}
\subsection{NOMA and ANOMA}

\begin{figure}[!ht b]
	\centering
	\includegraphics[width=2.5in]{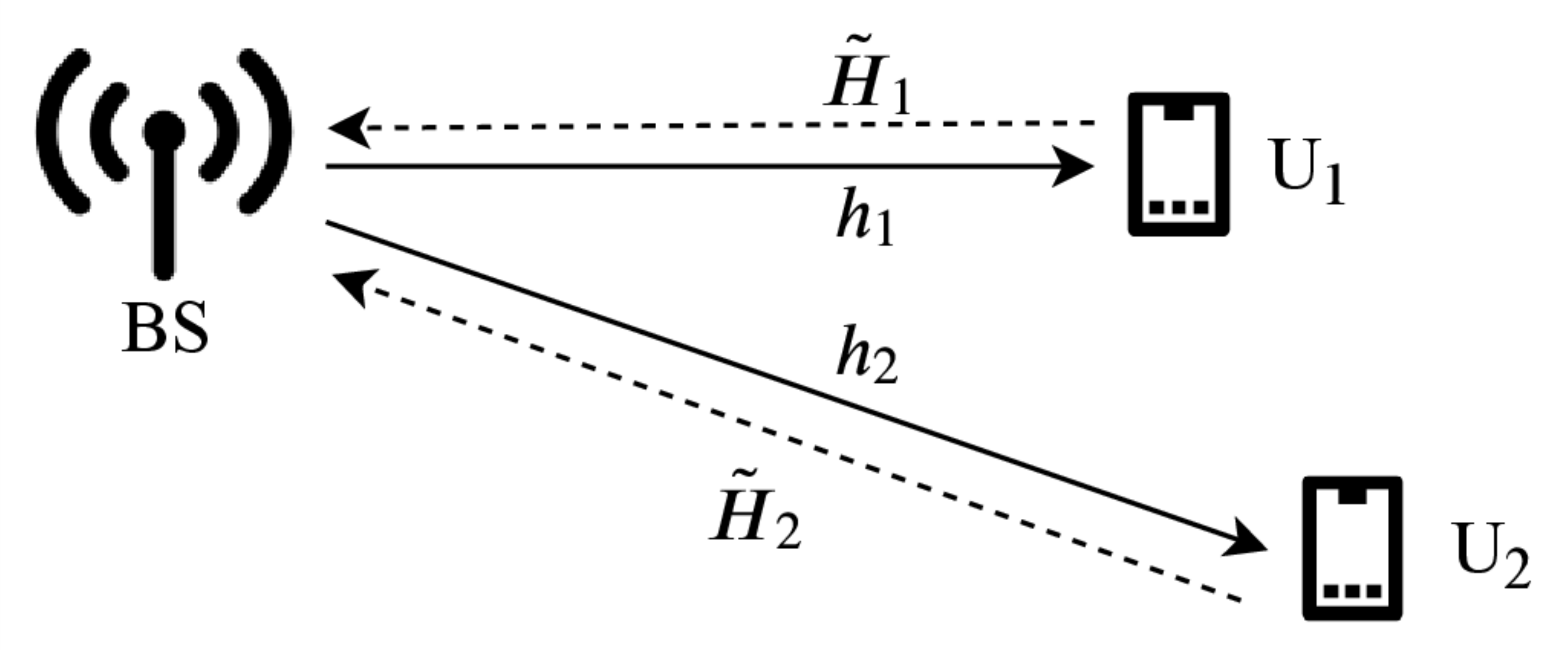}
	\caption{Downlink scenario with limited feedback.}
	\label{fig_scenario}
\end{figure}
\begin{figure}[!ht b]
	\centering
	\includegraphics[width=3.2in]{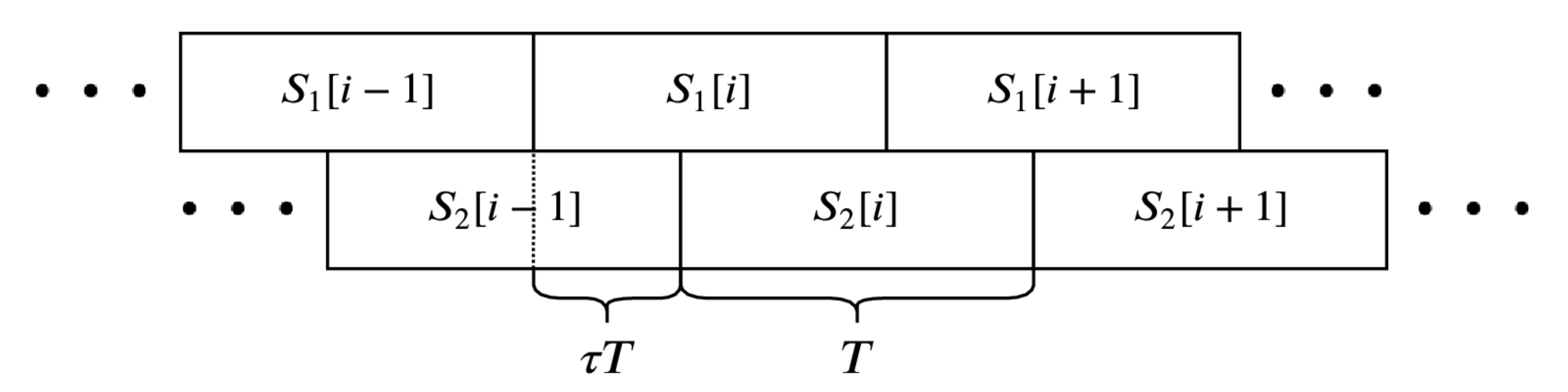}
	\caption{Illustration of the superimposed signal in ANOMA.}
	\label{fig_async_trans}
\end{figure}

In this letter, we consider a two-user downlink system shown in Fig.~\ref{fig_scenario} where the signals for Users~1 and 2 are superimposed and then transmitted by the BS. Both BS and users are equipped with a single antenna. As shown in Fig.~\ref{fig_async_trans}, a timing mismatch of $\tau T$ is intentionally added between the superimposed signals, where $0 < \tau < 1$ and $T$ is the symbol interval. Let $p(t)$ represent the pulse shape. The transmitted signal is given by $S_1[i]p(t-iT) + S_2[i]p(t-iT -\tau T)$ where $S_u[i]$, $u=1,2$, is the $i$th symbol transmitted to User~$u$. We assume that users know $\tau$ perfectly since it can be decided a priori, for example, to be the asymptotically optimal value of $\tau = 0.5$~\cite{zou2019analysis,zou2019cooperative}. The block fading channel model is employed. At the receiver, the composite signal is sampled at $iT$ and $(i+\tau)T$ after matched filtering. This sampling method is called ``oversampling'' and the details have already been presented in~\cite{zou2019analysis,zou2019cooperative}. For the sake of brevity, we omit it in this work. If $\tau = 0$, ANOMA degrades to NOMA and the received samples at $iT$ and $(i+\tau)T$ are identical. The number of received samples is doubled in ANOMA ($\tau \ne 0$) compared with NOMA, which then results in sampling diversity~\cite{zou2019analysis,zou2019cooperative,ganji2019improving}.

We assume that Users~1 and 2 denote the strong and weak users, respectively, i.e., $|h_1|^2 > |h_2|^2$. User~1 employs the block-wise SIC~\cite{zou2019cooperative}, i.e., first detects the block of the signal for User~2, removes it, and then detects its own signal. As shown in \cite{zou2019cooperative}, for a relatively large block length and the rectangular pulse shape, the rates of Users~1 and 2 with perfect CSI are given by
\begin{align}\label{eq_ANOMA_R1}
    R_{\mathrm{strong}}(H_1) = \log_2\left(1 \!+\! \alpha PH_1\right),
\end{align}

\noindent and \eqref{eq_ANOMA_R2} at the bottom of this page, $H_i = |h_i|^2$ is the channel gain of User~$i$, $Q = 2\tau(1-\tau)$, $P$ is the total transmit power of BS, $\alpha\in (0, 1)$ is the power coefficient for the strong user (User~1 in this case), i.e., the powers allocated to Users~1 and 2 are $\alpha P$ and $(1-\alpha)P$, respectively. Note that NOMA can be considered as a special case of ANOMA, simply by setting $\tau = 0$ in \eqref{eq_ANOMA_R2}. It has been shown that $R_{\mathrm{weak}}$ in ANOMA is higher than that in NOMA while $R_{\mathrm{strong}}$ is the same for ANOMA and NOMA~\cite{zou2019cooperative}.

\begin{figure*}[!b]
\begin{align}\label{eq_ANOMA_R2}
    R_{\mathrm{weak}}(H_2) = \log_2\left(\frac{1 + PH_2 + \alpha(1-\alpha)P^2H_2^2Q + \sqrt{[1 + PH_2 + \alpha(1-\alpha)P^2H_2^2Q]^2 - \alpha^2(1-\alpha)^2P^4H_2^4Q^2}}{2(1 + \alpha PH_2)}\right).
\end{align}
\end{figure*}

\subsection{Limited Feedback}
\begin{figure}[!ht b]
	\centering
	\includegraphics[width=3in]{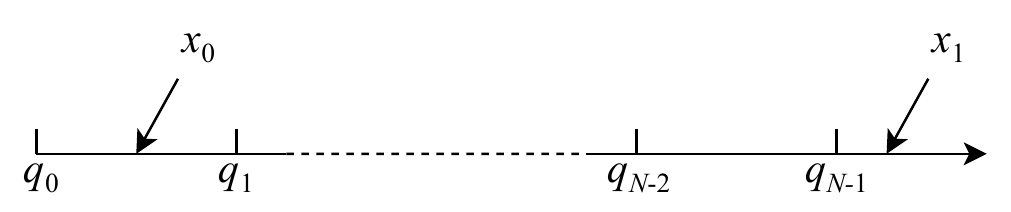}
	\caption{Illustration of a scalar quantizer.}
	\label{fig_quantizer}
\end{figure}

In this letter, we assume that CSI is perfectly estimated by users and fed back to the BS via an error and delay-free link. For example, in Fig.~\ref{fig_scenario}, User~1 knows $h_1$ and quantizes the channel gain $H_1$ as $\tilde{H}_1 = q(|h_1|^2)$ using a scalar quantizer $q$, and then sends it back to the BS. The BS determines the power coefficient $\alpha$ and the order of SIC according to the feedback. If $\tilde{H}_1 > \tilde{H}_2$ ($\tilde{H}_1 < \tilde{H}_2$), User~1 (User~2) is notified to conduct SIC. If $\tilde{H}_1 = \tilde{H}_2$, BS can randomly assign one user to utilize SIC. Without loss of generality, we assume that User~1 will be notified to conduct SIC if $\tilde{H}_1 = \tilde{H}_2$.

As shown in Fig.~\ref{fig_quantizer}, we employ a scalar quantizer with $N = 2^b$ quantization levels, $q_0, \cdots, q_{N-1}$, where $b$ is the number of bits used in a quantization codeword. $b$ is also defined as the feedback rate per user. The quantized value for a given $x$ is calculated by
\begin{align}\label{eq_quantizer}
    q(x) = \left\{  
    \begin{array}{lr}  
    q_i, & q_i \le x < q_{i+1},\ i=0,\cdots,N-2,\\
    q_{N-1}, & x \ge q_{N-1}.    
        \end{array}  \right.
\end{align}

For example, in Fig.~\ref{fig_quantizer}, $x_0$ and $x_1$ are quantized as $q_0$ and $q_{N-1}$, respectively. In this work, the quantizer is used to quantize the positive channel gain. Thus, we set $q_0 = 0$.

The BS transmits the signals to users based on the rates which are calculated according to the quantized channel gains. For example, if $\tilde{H}_1 > \tilde{H}_2$, the BS transmits to User~1 with the rate of $R_{\mathrm{strong}}(\tilde{H}_1)$ while the actual channel capacity is $R_{\mathrm{strong}}(H_1)$. Note that the proposed quantizer is designed to satisfy $q(x) \le q$ in order to avoid outage, i.e, to guarantee that the transmission rates do not exceed the channel capacities. It is trivial to derive that $R_{\mathrm{strong}}(\tilde{H}_1) \le R_{\mathrm{strong}}(H_1)$ if $\tilde{H}_1 \le H_1$ and $R_{\mathrm{weak}}(\tilde{H}_2) \le R_{\mathrm{weak}}(H_2)$ if $\tilde{H}_2 \le H_2$, which indicates that the quantizer in~\eqref{eq_quantizer} avoids outage.

\section{Power Allocation}
To attain fairness among users, we employ the max-min criterion for the power allocation, i.e., the power is allocated to users such that the minimum rate is maximized. The optimal power coefficient $\alpha^*$ is given by
\begin{align}\label{eq_maxmin_problem}
    \alpha^* = \mathop{\arg}_{\alpha} \max\min \left\{R_{\mathrm{strong}}, R_{\mathrm{weak}}\right\}.
\end{align}

According to \eqref{eq_ANOMA_R1} and \eqref{eq_ANOMA_R2}, it is trivial to show that $R_{\mathrm{strong}}$ is an increasing function of $\alpha$ and $R_{\mathrm{weak}}$ is a decreasing function of $\alpha$. Intuitively, if User~1 is the strong user, by increasing $\alpha$ (i.e., allocating more power to User~1 and less power to User~2), $R_{\mathrm{strong}}$ increases and $R_{\mathrm{weak}}$ decreases, and vice versa. As a result, the optimal power coefficient $\alpha^*$ can be obtained by solving $R_{\mathrm{strong}} = R_{\mathrm{weak}}$. According to \cite{liu2017downlink}, the optimal $\alpha$ in NOMA is given by
\begin{align}\label{eq_NOMA_maxmin_rate}
    \alpha_\mathrm{N}^* \!=\! \frac{2\tilde{H}_{\min}}{\sqrt{(\tilde{H}_1\!+\!\tilde{H}_2)^2 \!+\! 4P\tilde{H}_1\tilde{H}_2\tilde{H}_{\min}} \!+\! \tilde{H}_1 \!+\! \tilde{H}_2},
\end{align}
where $\tilde{H}_{\min} = \min\left\{\tilde{H}_1, \tilde{H}_2\right\}$ is defined to incorporate the rate expressions for both $\tilde{H}_1 \ge \tilde{H}_2$ and $\tilde{H}_1 < \tilde{H}_2$. If User~1 employs SIC, the max-min rate of NOMA is given by $\log\left(1 + \alpha_\mathrm{N}^*P\tilde{H_1}\right)$ according to~\eqref{eq_ANOMA_R1}. If User~2 employs SIC, the max-min rate is given by $\log\left(1 + \alpha_\mathrm{N}^*P\tilde{H_2}\right)$. To summarize, the max-min rate of NOMA is expressed by
$R_{\mathrm{N}}^*(\tilde{H}_1, \tilde{H}_2) = \log\left(1 + \alpha_\mathrm{N}^*P\tilde{H}_{\max}\right)$ where $\tilde{H}_{\max} = \max\left\{\tilde{H}_1, \tilde{H}_2\right\}$.

The optimal power coefficient for ANOMA, $\alpha_{\mathrm{A}}^*$, is presented in the following theorem. 
\begin{Theorem}\label{theorem_NOMA_vs_ANOMA}
The optimal power coefficients of NOMA and ANOMA, $\alpha_{\mathrm{N}}^*$ and $\alpha_{\mathrm{A}}^*$, respectively, satisfy the following inequality
\begin{align}\label{eq_theorem}
    \alpha_\mathrm{N}^* \le \alpha_\mathrm{A}^*(0.5)\le
    \alpha_\mathrm{A}^* \le
    \alpha_\mathrm{A}^*(1),
\end{align} 
where the equal sign is achieved when $\tau = 0$. The expression for $\alpha_\mathrm{A}^*(z)$ is given by~\eqref{eq_ANOMA_maxmin_rate} on the top of this page. 
\begin{figure*}[htb]
\begin{align}\label{eq_ANOMA_maxmin_rate}
    \alpha_{\mathrm{A}}^*(z) = \frac{2\tilde{H}_{\min}}{\sqrt{(\tilde{H}_1\!+\!\tilde{H}_2 - z P\tilde{H}_{\min}^2Q)^2 \!+\! 4\tilde{H}_{\min}(P\tilde{H}_1\tilde{H}_2 + z P\tilde{H}_{\min}^2Q)} \!+\! \tilde{H}_1 \!+\! \tilde{H}_2 - z P\tilde{H}_{\min}^2Q}.
\end{align}
\end{figure*}
\end{Theorem}
\begin{IEEEproof}
See Appendix \ref{appendix1}.
\end{IEEEproof}

Since the general expression for $\alpha_\mathrm{A}^*$ is intractable for further analysis, $\alpha_\mathrm{A}^*(0.5)$ and $\alpha_\mathrm{A}^*(1)$ provide the simple lower and upper bounds of $\alpha_\mathrm{A}^*$, respectively. Compared with \eqref{eq_NOMA_maxmin_rate}, \eqref{eq_ANOMA_maxmin_rate} introduces an extra term $zP\tilde{H}_{\min}^2Q$.
By setting $\tau = 0$, $Q = 0$ and then \eqref{eq_ANOMA_maxmin_rate} coincides with \eqref{eq_NOMA_maxmin_rate}. The max-min rate of ANOMA is given by $R_{\mathrm{A}}^*(\tilde{H}_1, \tilde{H}_2) = \log\left(1 + \alpha_{\mathrm{A}}^*P\tilde{H}_{\max}\right) \in \left[\log\left(1 + \alpha_{\mathrm{A}}^*(0.5)P\tilde{H}_{\max}\right), \log\left(1 + \alpha_{\mathrm{A}}^*(1)P\tilde{H}_{\max}\right)\right]$.

The average max-min rate is expressed by
\begin{align}\label{eq_ER}
    \mathbb{E}\left[R^*\right] =& \sum_{i=0}^{N_1-1}\sum_{j=0}^{N_2-1}\int_{q_{i,1}}^{q_{i+1,1}}\int_{q_{j,2}}^{q_{j+1,2}}R^*(q_{i,1}, q_{j,2})\notag\\
    &\cdot f_1(H_1)f_2(H_2)\mathrm{d}H_1\mathrm{d}H_2,
\end{align}
where $f_i(H_i)$ is the distribution function of User~$i$'s channel gain, $q_{j,i}$ represents the $j$th quantization level of the quantizer used by User~$i$, and $R^*$ can be $R^*_\mathrm{N}$ or $R^*_\mathrm{A}$. Note that the average max-min rate for the limited feedback is upper bounded by that for the full-CSI case, i.e.,
\begin{align}\label{eq_asymptotic_val}
    \mathbb{E}\left[R^*\right] \!<\! \overline{R}^*\! \stackrel{\triangle}{=}\!\! \int_{0}^{\infty}\!\!\!\!\int_{0}^{\infty}\!\!\!\!R^*(H_1,\! H_2)f_1(\!H_1\!)f_2(H_2)\mathrm{d}H_1\mathrm{d}H_2.
\end{align}
Let us define the quantization distortion as $D\left[R^*\right] = \overline{R}^* - \mathbb{E}\left[R^*\right]$.

\begin{Corollary}\label{corollary}
ANOMA can achieve the average max-min rate of NOMA with a lower feedback rate.
\end{Corollary}
\begin{IEEEproof}
Let us define $q$ and $q'$ as two quantizers which can be given by~\eqref{eq_quantizer} but with different quantization levels. According to Theorem~\ref{theorem_NOMA_vs_ANOMA}, ANOMA achieves a higher max-min rate compared with NOMA. Thus, $\overline{R}^*_{\mathrm{N}} < \overline{R}^*_{\mathrm{A}}$ and $\mathbb{E}\left[R^*_{\mathrm{N}}\right] < \mathbb{E}\left[R^*_{\mathrm{A}}\right]$ by using the quantizer $q$. The quantizer $q'$ is designed such that the ANOMA using $q'$ achieves the same average max-min rate as the NOMA using $q$, i.e., $\mathbb{E}\left[R^*_{\mathrm{A}}\right]' = \mathbb{E}\left[R^*_{\mathrm{N}}\right] < \mathbb{E}\left[R^*_{\mathrm{A}}\right]$. For ANOMA, using the quantizer $q'$ results in a higher distortion compared with using $q$, i.e., $D\left[R^*_\mathrm{A}\right]' = \overline{R}^*_{\mathrm{A}} - \mathbb{E}\left[R^*_{\mathrm{A}}\right]' > D\left[R^*_\mathrm{A}\right] = \overline{R}^*_{\mathrm{A}} - \mathbb{E}\left[R^*_{\mathrm{A}}\right]$. According to the rate-distortion theory, there is a trade-off between the distortion and the quantization rate (equivalent to the feedback rate in this work). A lower distortion can be achieved by using a quantizer with a higher feedback rate and vice versa. As a result, the quantizer $q'$ can have a lower feedback rate compared with $q$. ANOMA using $q'$ can achieve a lower feedback rate while keeping the same average max-min rate as NOMA using $q$. The proof is complete.
\end{IEEEproof}

\section{Scalar Quantizer Optimization}
The scalar quantizer shown in~\eqref{eq_quantizer} is completely characterized by the quantization levels. Our goal is to optimize the quantization levels to maximize the average max-min rate $\mathbb{E}\left[R^*\right]$, i.e.,
\begin{align}\label{eq_optimization_problem}
    [\mathbf{q}_1^*, \mathbf{q}_2^*] =& \mathop{\arg}_{[\mathbf{q_1}, \mathbf{q_2}]}\max \mathbb{E}\left[R^*\right], \notag\\
    &s.t.\ q_{0,i}<q_{1,i}<\cdots < q_{N_i-1, i},\ i=1\ \mathrm{or}\ 2.
\end{align}
Since $\overline{R}^*$ is not a function of quantization levels, \eqref{eq_optimization_problem} is equivalent to minimizing the distortion $D[R^*]$. In what follows, we propose a gradient descent algorithm to optimize the quantization levels. Let $\mathbb{E}\left[R^*\right]_{i, j}$ denote the ($i,j$)th term of $\mathbb{E}\left[R^*\right]$ in~\eqref{eq_ER}, i.e., $\mathbb{E}\left[R^*\right]_{i,j} =R^*(q_{i,1}, q_{j,2})\int_{q_{i,1}}^{q_{i+1,1}}\!\!\int_{q_{j,2}}^{q_{j+1,2}}\!\!f_1(H_1)f_2(H_2)\mathrm{d}H_1\mathrm{d}H_2$. $\mathbb{E}\left[R^*\right]_{i, j}$ is a function of $q_{i,1}$, $q_{i+1,1}$, $q_{j,2}$, and $q_{j+1,2}$. The gradients of $\mathbb{E}\left[R^*\right]_{i, j}$ in terms of $q_{i,1}$ and $q_{i+1,1}$ are calculated by
\begin{align}
    \frac{\partial \mathbb{E}\left[R^*\right]_{i, j}}{\partial q_{i,1}} \!=\!& \left[\frac{\partial R^*(q_{i,1}, q_{j,2})}{\partial q_{i, 1}}\int_{q_{i,1}}^{q_{i+1,1}}f_1(H_1)\mathrm{d}H_1\right. \notag\\
    &\left.\!- R^*(q_{i,1}, q_{j,2})f_1(q_{i,1})\right]\!\int_{q_{j,2}}^{q_{j+1,\!2}}\!\!\!\!f_2(H_2)\mathrm{d}H_2,\\
    \frac{\partial \mathbb{E}\left[R^*\right]_{i, j}}{\partial q_{i+1,1}} \!=\!& R^*(q_{i,1}, q_{j,2})f_1(q_{i+1,1}) \!\!\int_{q_{j,2}}^{q_{j+1,2}}\!\!\!\!f_2(H_2)\mathrm{d}H_2,
\end{align}
respectively. Similarly, we can derive $\frac{\partial \mathbb{E}\left[R^*\right]_{i, j}}{\partial q_{j,2}}$ and $\frac{\partial \mathbb{E}\left[R^*\right]_{i, j}}{\partial q_{j+1,2}}$. Based on the gradients, we propose the quantizer optimization algorithm in Algorithm 1. At each iteration, $\mathbb{E}\left[R^*\right]$ does not decrease which is guaranteed by the gradient descent. Besides, $\mathbb{E}\left[R^*\right]$ is upper bounded by a constant as shown in \eqref{eq_asymptotic_val}. Hence, Algorithm~1 converges as the number of iterations increases.

\begin{algorithm}[!h]
\caption{Algorithm to optimize the scalar quantizer.}
 \begin{algorithmic}[1]
\State Initialize the step size $\Delta$, the maximum number of iterations $I_{\max}$, the number of quantization levels for User~1 $N_1$ and that for User~2 $N_2$, the iteration count $I = 0$.
\State Initialize the quantization levels $\mathbf{q}_{1} = [0, q_{1,1}, \cdots, q_{N_1-1, 1}, \infty]$, $\mathbf{q}_{2} = [0, q_{1,2}, \cdots, q_{N_2-1, 2}, \infty]$
\While{$I < I_{\max}$}{}
\State Initialize $E_R = 0$, $\mathbf{dq}_1 = [0, \cdots, 0]_{1\times (N_1+1)}$, $\mathbf{dq}_2 = [0, \cdots, 0]_{1\times (N_2+1)}$.
\For{$i=2,\cdots, N_1$}
\For{$j=2,\cdots, N_2$}
\State $E_R = E_R + R^*(q_{1}[i], q_{2}[j])\int_{q_1[i]}^{q_{1}[i+1]}\int_{q_{2}[j]}^{q_{2}[j+1]}$
    $\cdot \!f_1(H_1)f_2(H_2)\mathrm{d}H_1\mathrm{d}H_2$.
\State $\mathbf{dq}_1[i] = \mathbf{dq}_1[i] + \frac{\partial \mathbb{E}\left[R^*\right]_{i,j}}{\partial q_{i,1}}$.
\State $\mathbf{dq}_2[j] = \mathbf{dq}_2[j] + \frac{\partial \mathbb{E}\left[R^*\right]_{i,j}}{\partial q_{j,2}}$.
\If{$i \ne N_1$}
\State $\mathbf{dq}_1[i+1] = \mathbf{dq}_1[i+1] + \frac{\partial \mathbb{E}\left[R^*\right]_{i,j}}{\partial q_{i+1,1}}$. 
\EndIf
\If{$j \ne N_2$}
\State $\mathbf{dq}_2[j+1] = \mathbf{dq}_2[j+1] + \frac{\partial \mathbb{E}\left[R^*\right]_{i,j}}{\partial q_{j+1,2}}$.
\EndIf
\EndFor
\EndFor
\State $\mathbf{q}_1 = \mathbf{q}_1 + \Delta * \mathbf{dq}_1$, $\mathbf{q}_2 = \mathbf{q}_2 + \Delta * \mathbf{dq}_2$, $I = I + 1$.
\EndWhile
 \end{algorithmic}
 \end{algorithm}

The computation complexity of Algorithm~1 is $O(N_1N_2)$ where $N_1$ and $N_2$ are the number of quantization levels for Users~1 and 2, respectively. Furthermore, the maximum quantization level of the conventional uniform quantizer is set manually according to certain criterion. For example, in~\cite{liu2017downlink}, the maximum quantization level is determined by considering the quantization loss. The advantage of optimizing the quantization levels is that the maximum quantization level can also be optimized using Algorithm~1 with no manual intervention, which will be shown in the next section.

\begin{figure}[!ht b]
	\centering
	\includegraphics[width=3.1in]{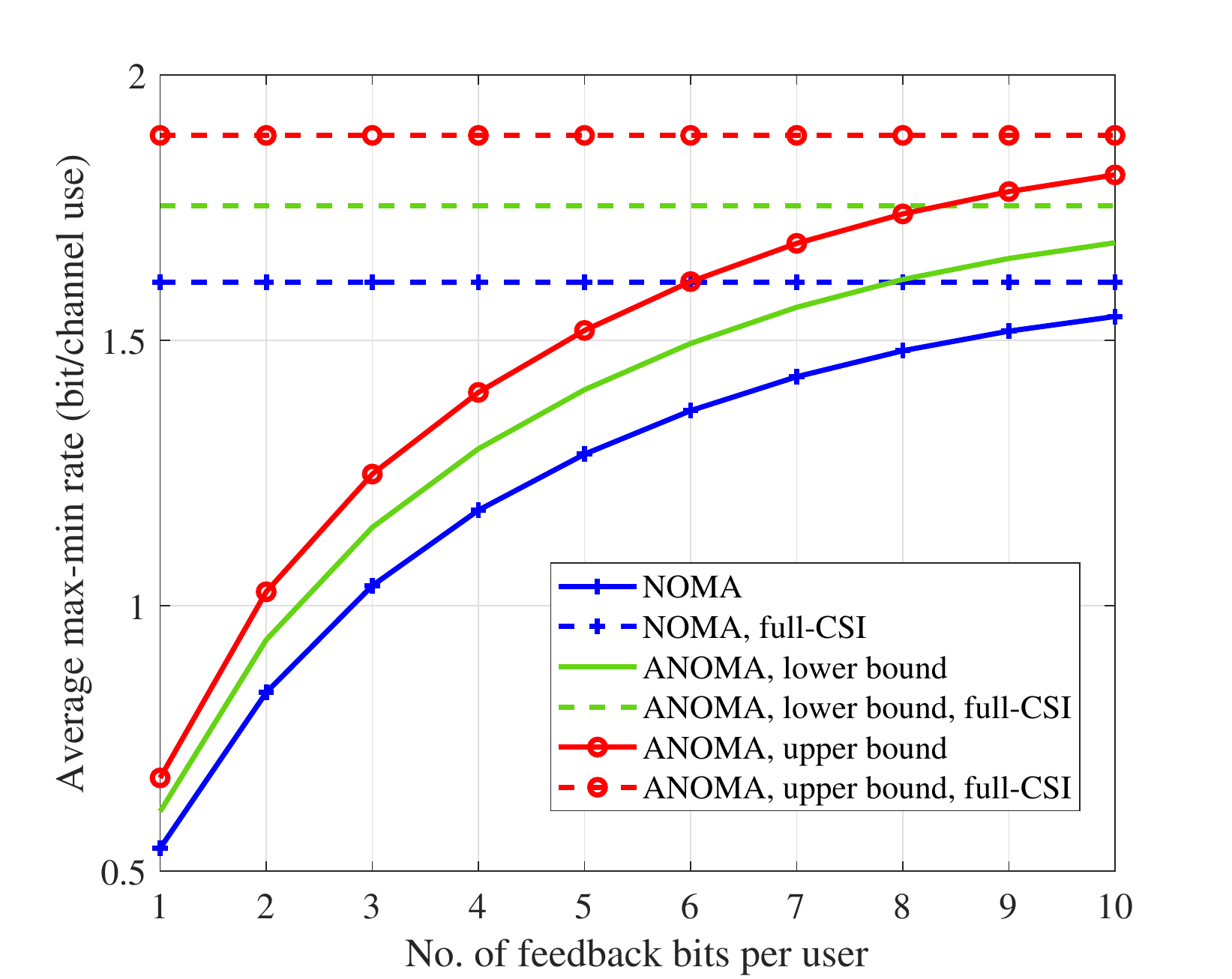}
	\caption{The average max-min rate vs. the number of feedback bits per channel for NOMA and ANOMA systems.}
	\label{fig_Erate_vs_feedback_rate}
\end{figure}
\section{Simulation Results}
In this section, we present simulation results for NOMA/ANOMA systems with limited feedback. In our simulations, we employ the complex Gaussian channel model. Therefore, the channel gain follows the exponential distribution. We assume that $H_1 \sim \mathrm{Exp}(0.5)$ and $H_2 \sim \mathrm{Exp}(1)$. We set the total transmit power of the BS $P = 10$ and $\tau = 0.5$ since it has been proved in \cite{zou2019analysis,zou2019cooperative} that $\tau = 0.5$ is the asymptotically optimal value to maximize the user throughput. For comparison, we employ the uniform quantizer proposed in~\cite{liu2017downlink} where the maximum quantization level $L$ is derived by solving $L = \frac{1}{\lambda\Delta}\log\left(\frac{1}{\Delta}\right)$, $\lambda$ is the parameter of the exponential distribution, and $\Delta$ is the quantization bin width. 
\begin{figure}[!t b]
	\centering
	\includegraphics[width=3.1in]{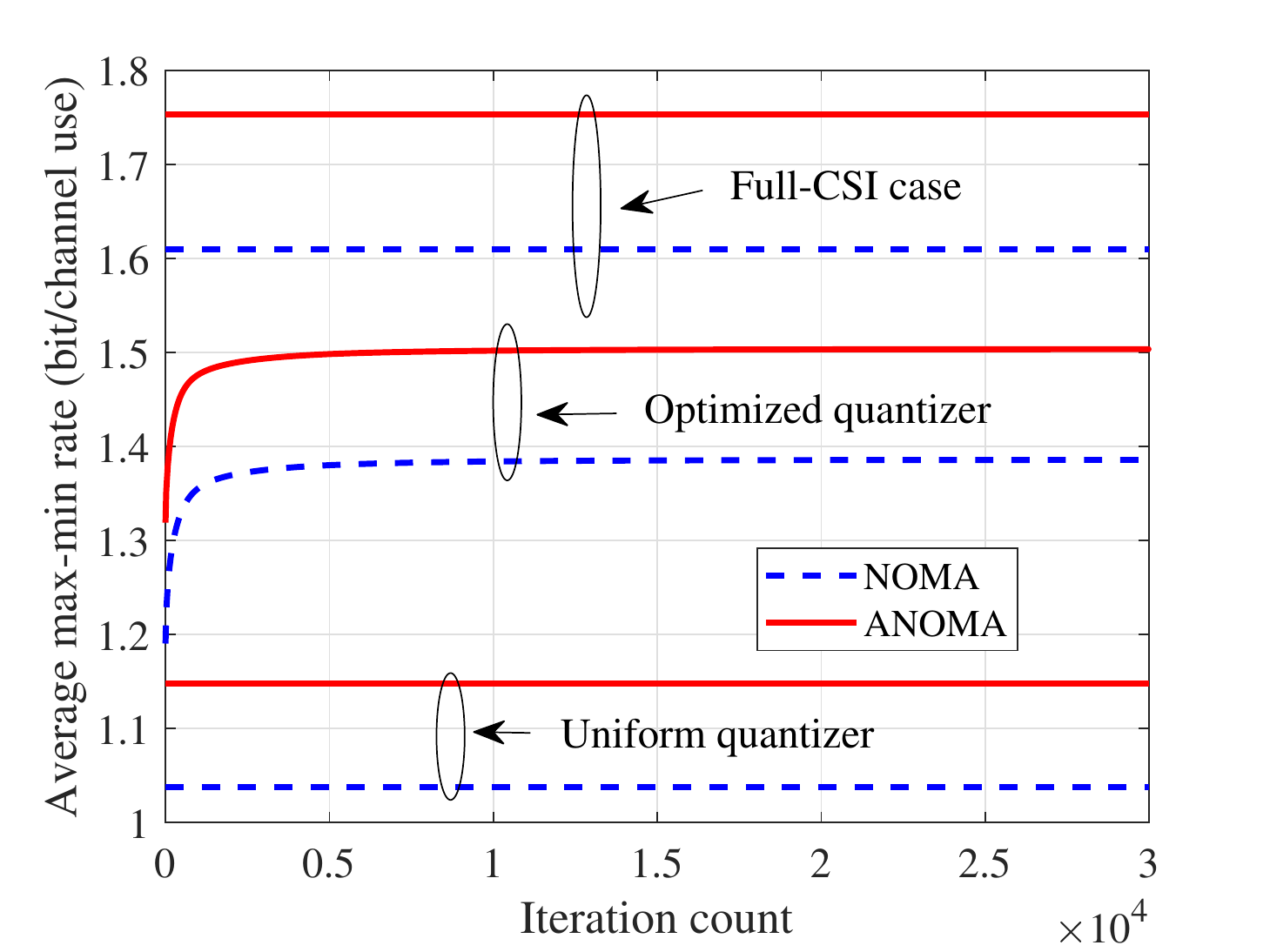}
	\caption{The average max-min rate vs. the iteration count for NOMA and ANOMA systems.}
	\label{fig_Erate_vs_iteration_count}
\end{figure}
\begin{figure}[!t b]
	\centering
	\includegraphics[width=3.1in]{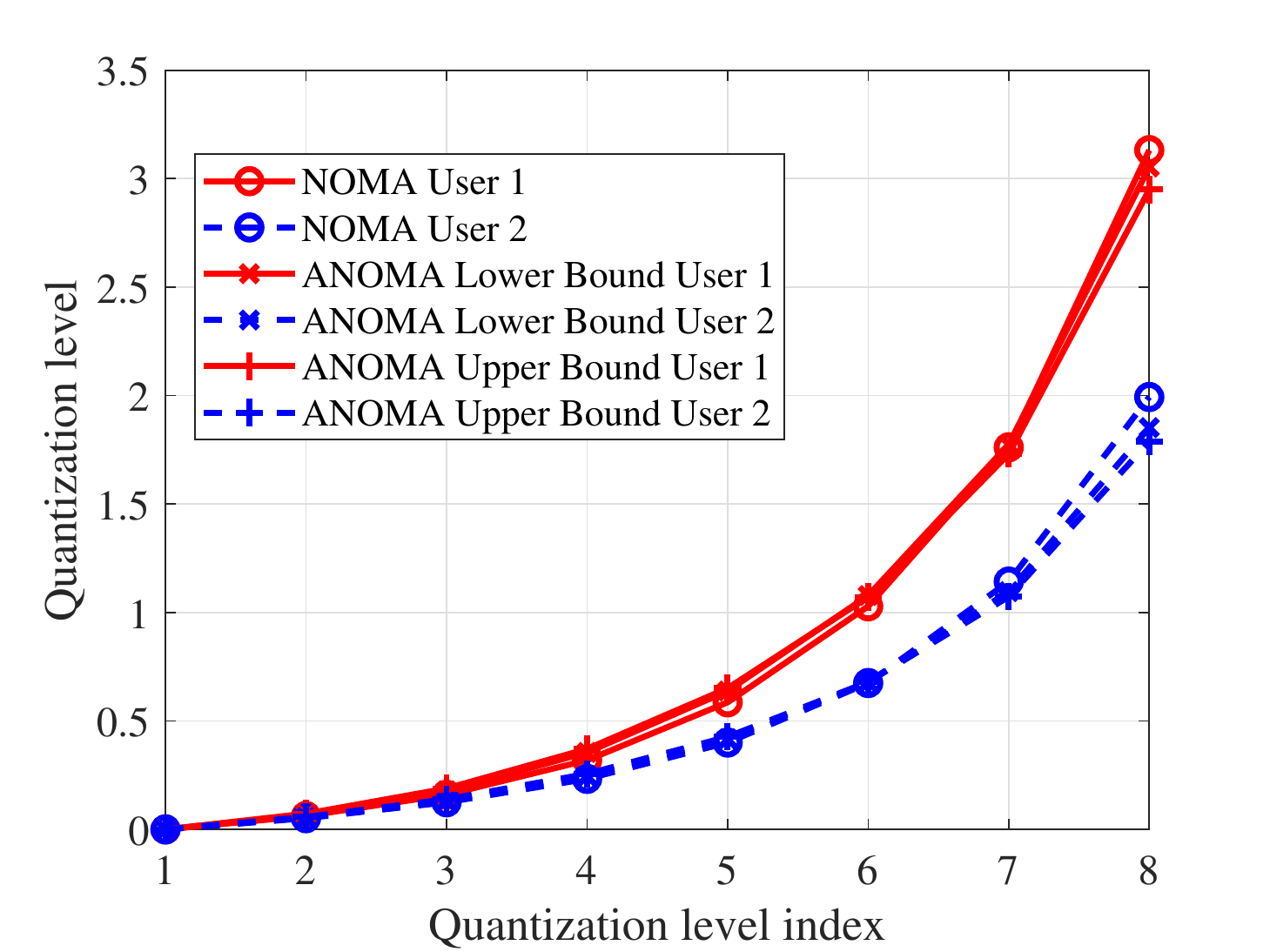}
	\caption{The quantization levels for NOMA and ANOMA systems when 3 bits are used to quantize the channel gain.}
	\label{fig_channel_gain_vs_level}
\end{figure}

Fig.~\ref{fig_Erate_vs_feedback_rate} shows how the average max-min rate changes as a function of the number of feedback bits per user using the uniform quantizer. As the number of feedback bits increases, the average max-min rate converges to the full-CSI rate calculated by \eqref{eq_asymptotic_val}. Besides, using the same number of feedback bits, the upper bound of the max-min rate for ANOMA is always higher than its lower bound, which is then higher than that for NOMA. Equivalently, to achieve the same or even higher average max-min rate, ANOMA needs less feedback bits compared with NOMA, which verifies Corollary~\ref{corollary}.

Fig.~\ref{fig_Erate_vs_iteration_count} shows how the average max-min rate in NOMA and ANOMA systems changes as the quantizer optimization algorithm runs. We employ the 3-bit quantizer for each channel and the lower bound of the max-min rate for ANOMA as an example. As the number of iterations increases, the average max-min rate converges. For the full-CSI case, the optimized quantizer, and the uniform quantizer, the average max-min rate for ANOMA is always higher than that for NOMA. 

Fig.~\ref{fig_channel_gain_vs_level} presents the optimal quantization levels when 3 bits (i.e., $2^3 = 8$ quantization levels) are used. First, it is shown that the optimal quantization levels are non-uniform. The range of the quantization levels for User~1 with a larger mean channel gain is wider than that of User~2. Furthermore, the quantization levels optimized using the upper and lower bounds of the max-min rate for ANOMA are very similar to each other and also close to those for NOMA.

\section{Conclusion}
In this letter, we considered a two-user downlink ANOMA system with limited feedback. We reveal the advantage of ANOMA over NOMA in terms of the feedback rate. Furthermore, the quantizer optimization method is proposed for both ANOMA and NOMA to further improve the average max-min rate. Considering more than two users in the ANOMA systems is an interesting topic for the future work.
\vspace{-3mm}
\appendices
\section{Proof of Theorem~\ref{theorem_NOMA_vs_ANOMA}}\label{appendix1}
\begin{IEEEproof}
When $\tau = 0$, $Q = 0$ which then results in $\alpha_\mathrm{N}^* = \alpha_\mathrm{A}^*(x)$ for any finite $x$. For $\tau \ne 0$,
we first prove Theorem~\ref{theorem_NOMA_vs_ANOMA} for the case of $\tilde{H}_1 \ge \tilde{H}_2$. The case of $\tilde{H}_1 < \tilde{H}_2$ SIC will be discussed later. By setting $R_{\mathrm{strong}}(\tilde{H}_1) = R_{\mathrm{weak}}(\tilde{H}_2)$, we obtain
\begin{align}\label{eq_quartic_eqation}
    2&[1+\alpha^*_{\mathrm{A}} P(\tilde{H}_1 + \tilde{H}_2) + (\alpha^{*}_{\mathrm{A}})^{2}P^2\tilde{H}_1\tilde{H}_2] \notag\\
    =& \sqrt{[1 \!+\! P\tilde{H}_2 \!+\! \alpha^*_{\mathrm{A}}(1\!-\!\alpha^*_{\mathrm{A}})P^2\tilde{H}_2^2Q]^2 \!-\! [\alpha^*_{\mathrm{A}}(1\!-\!\alpha^*_{\mathrm{A}})]^2P^4\tilde{H}_2^4Q^2}\notag\\ 
    &+ 1 + P\tilde{H}_2 + \alpha^*_{\mathrm{A}}(1-\alpha^*_{\mathrm{A}})P^2\tilde{H}_2^2Q.
\end{align}
By cancelling out the square root, \eqref{eq_quartic_eqation} becomes a quartic equation. The optimal power coefficient $\alpha^*_{\mathrm{A}}$ is one root of the quartic equation which can be given by the general formula for quartic roots. However, it is intractable for further analysis. Therefore, we derive the upper and lower bounds to approximate the actual value of $\alpha^*_{\mathrm{A}}$. 

For the upper bound, since $[1 \!+\! P\tilde{H}_2 \!+\! \alpha^*_{\mathrm{A}}(1\!-\!\alpha^*_{\mathrm{A}})P^2\tilde{H}_2^2Q]^2 \!-\! (\alpha^*_{\mathrm{A}})^2(1\!-\!\alpha^*_{\mathrm{A}})^2P^4\tilde{H}_2^4Q^2 < [1 \!+\! P\tilde{H}_2 \!+\! \alpha^*_{\mathrm{A}}(1\!-\!\alpha^*_{\mathrm{A}})P^2\tilde{H}_2^2Q]^2$, \eqref{eq_quartic_eqation} becomes
\begin{align}
    1\!+\!\alpha^*_{\mathrm{A}} P(\tilde{H}_1 \!+\! \tilde{H}_2) \!&+\! (\alpha^*_{\mathrm{A}})^2P^2\tilde{H}_1\tilde{H}_2\notag\\
    &<\! 1 \!+ \!P\tilde{H}_2 \!+\! \alpha^*_{\mathrm{A}}\!(1\!-\!\alpha^*_{\mathrm{A}})P^2\tilde{H}_2^2Q,
\end{align}
which results in
\begin{align}\label{eq_alpha_low_bound}
    &\alpha^*_{\mathrm{A}} \!<\!  \alpha^*_U\! \stackrel{\triangle}{=}\! 2\tilde{H}_2/\left[\tilde{H}_1 \!+\! \tilde{H}_2 - P\tilde{H}_{2}^2Q\right.\notag\\
    &\left.\!+\! \sqrt{(\tilde{H}_1\!+\!\tilde{H}_2 \!-\! P\tilde{H}_{2}^2Q)^2 \!+\! 4\tilde{H}_{2}(P\tilde{H}_1\tilde{H}_2 \!+\! P\tilde{H}_{2}^2Q)} \right].
\end{align}

For the lower bound, as $[1 \!+\! P\tilde{H}_2 \!+\! \alpha^*_{\mathrm{A}}(1\!-\!\alpha^*_{\mathrm{A}})P^2\tilde{H}_2^2Q]^2 \!-\! (\alpha^*_{\mathrm{A}})^2(1\!-\!\alpha^*_{\mathrm{A}})^2P^4\tilde{H}_2^4Q^2 > [1 \!+\! P\tilde{H}_2]^2$,
\begin{align}
    2[1\!+\!\alpha^*_{\mathrm{A}} P(\tilde{H}_1 \!&+\! \tilde{H}_2) \!+\! (\alpha^*_{\mathrm{A}})^2P^2\tilde{H}_1\tilde{H}_2]\notag\\
    &> 2(1 \!+ \!P\tilde{H}_2)
    +\! \alpha^*_{\mathrm{A}}(1\!-\!\alpha^*_{\mathrm{A}})P^2\tilde{H}_2^2Q.
\end{align}

Then, we obtain
\begin{align}\label{eq_alpha_up_bound}
    &\alpha^*_{\mathrm{A}} \!>  \alpha^*_L \stackrel{\triangle}{=} 2\tilde{H}_2/\left[\tilde{H}_1 \!+\! \tilde{H}_2 - \frac{P\!\tilde{H}_{2}^2Q}{2}\right.\notag\\
    &\left.\!+\! \sqrt{\!\left(\!\tilde{H}_1\!+\!\tilde{H}_2 \!-\! \frac{P\!\tilde{H}_{2}^2Q}{2}\! \right)^{\!2}\!\!\! \!+\! 4\tilde{H}_{2}\!\left(P\tilde{H}_1\tilde{H}_2 \!+\! \frac{P\tilde{H}_{2}^2Q}{2}\!\right)} \!\right].
\end{align}

If $\tilde{H}_1 < \tilde{H}_2$, i.e., User~2 employs SIC, we can also derive the expressions for $\alpha^*_{L}$ and $\alpha^*_{U}$ by setting $R_{\mathrm{strong}}(\tilde{H}_2) = R_{\mathrm{weak}}(\tilde{H}_1)$. In fact, $\alpha^*_{L}$ and $\alpha^*_{U}$ for $\tilde{H}_1 < \tilde{H}_2$ are given by simply switching $\tilde{H}_1$ and $\tilde{H}_2$ in \eqref{eq_alpha_low_bound} and \eqref{eq_alpha_up_bound}. Both $\alpha^*_L$ and $\alpha^*_U$ can be incorporated into a general expression in~\eqref{eq_ANOMA_maxmin_rate} by introducing a parameter $z$, i.e. $\alpha_{L}^* = \alpha_{\mathrm{A}}^*(0.5)$ and $\alpha_{U}^* = \alpha_{\mathrm{A}}^*(1)$.



To show the inequality in \eqref{eq_theorem}, let us define $g(x) = \sqrt{(\tilde{H}_1\!+\!\tilde{H}_2 - xP\tilde{H}_{\min}^2Q)^2 \!+\! 4\tilde{H}_{\min}(P\tilde{H}_1\tilde{H}_2 + xP\tilde{H}_{\min}^2Q)} + \tilde{H}_1 \!+\! \tilde{H}_2 - xP\tilde{H}_{\min}^2Q$ which is the denominator in \eqref{eq_ANOMA_maxmin_rate}. Then,
\begin{align}
    &\frac{\partial g(x)}{\partial x} = -P\tilde{H}_{\min}^2Q\left(1+\right.\notag\\
    &\left.{\frac{\tilde{H}_1 + \tilde{H}_2 - \tilde{H}_{\min}(2+xP\tilde{H}_{\min}Q)}{\sqrt{\!(\tilde{H}_1\!+\!\tilde{H}_2 \!-\! xP\tilde{H}_{\min}^2Q)^2 \!+\! 4\tilde{H}_{\min}(P\tilde{H}_1\tilde{H}_2 \!+\! xP\tilde{H}_{\min}^2Q)}}}\!\right).\notag
\end{align}

If $\tilde{H}_1 + \tilde{H}_2 - \tilde{H}_{\min}(2+P\tilde{H}_{\min}Qx) > 0$, it is obvious that $\frac{\partial g(x)}{\partial x} < 0$. Otherwise,
\begin{align}
    &-\tilde{H}_1 -\tilde{H}_2 + \tilde{H}_{\min} < 0 < P\tilde{H}_1\tilde{H}_2\notag\\
    \Longrightarrow &4\tilde{H}_{\min}(-\tilde{H}_1 -\tilde{H}_2 + \tilde{H}_{\min} + xP\tilde{H}_{\min}^2Q)\notag\\
    &< 4\tilde{H}_{\min}(P\tilde{H}_1\tilde{H}_2 + xP\tilde{H}_{\min}^2Q) \notag\\
    \Longrightarrow &|\tilde{H}_1 + \tilde{H}_2 - \tilde{H}_{\min}(2+xP\tilde{H}_{\min}Q)|\notag\\ 
    &< \!\!\!\sqrt{\!(\tilde{H}_1\!+\!\tilde{H}_2 \!-\! xP\tilde{H}_{\min}^2Q)^2 \!+\! 4P\tilde{H}_{\min}(\tilde{H}_1\tilde{H}_2 \!+\! x\tilde{H}_{\min}^2Q)},\notag
\end{align}
which also results in $\frac{\partial g(x)}{\partial x} < 0$. Therefore, $\alpha^*_{\mathrm{A}}(x)$ increases with $x$. Thus, $\alpha_{\mathrm{N}}^* = \alpha_{\mathrm{A}}^*(0) < \alpha_{\mathrm{A}}^*(0.5) < \alpha_{\mathrm{A}}^* < \alpha_{\mathrm{A}}^*(1)$. The proof is complete.
\end{IEEEproof}

{\small
\bibliographystyle{IEEEtran}
\bibliography{IEEEabrv,IEEEexample}

\begin{thebibliography}{1}
\providecommand{\url}[1]{#1}
\csname url@samestyle\endcsname
\providecommand{\newblock}{\relax}
\providecommand{\bibinfo}[2]{#2}
\providecommand{\BIBentrySTDinterwordspacing}{\spaceskip=0pt\relax}
\providecommand{\BIBentryALTinterwordstretchfactor}{4}
\providecommand{\BIBentryALTinterwordspacing}{\spaceskip=\fontdimen2\font plus
\BIBentryALTinterwordstretchfactor\fontdimen3\font minus
  \fontdimen4\font\relax}
\providecommand{\BIBforeignlanguage}[2]{{%
\expandafter\ifx\csname l@#1\endcsname\relax
\typeout{** WARNING: IEEEtran.bst: No hyphenation pattern has been}%
\typeout{** loaded for the language `#1'. Using the pattern for}%
\typeout{** the default language instead.}%
\else
\language=\csname l@#1\endcsname
\fi
#2}}
\providecommand{\BIBdecl}{\relax}
\BIBdecl

\bibitem{zou2019TCNOMA}
X.~Zou, M.~Ganji, and H.~Jafarkhani, ``Trellis-coded non-orthogonal multiple
  access,'' \emph{{IEEE} Wireless Commun. Lett.}, vol.~9, no.~4, pp. 538--542,
  Apr. 2020.

\bibitem{zou2019analysis}
X.~Zou, B.~He, and H.~Jafarkhani, ``An analysis of two-user uplink asynchronous
  non-orthogonal multiple access systems,'' \emph{{IEEE} Trans. Wireless
  Commun.}, vol.~18, no.~2, pp. 1404--1418, Feb. 2019.

\bibitem{zou2019cooperative}
X.~Zou, M.~Ganji, and H.~Jafarkhani, ``Cooperative asynchronous non-orthogonal
  multiple access with power minimization under {QoS} constraints,''
  \emph{{IEEE} Trans. Wireless Commun.}, vol.~19, no.~3, pp. 1503--1518, Mar.
  2020.

\bibitem{ganji2019improving}
M.~Ganji and H.~Jafarkhani, ``Improving {NOMA} multi-carrier systems with
  intentional frequency offsets,'' \emph{{IEEE} Wireless Commun. Lett.},
  vol.~8, no.~4, pp. 1060--1063, Aug. 2019.

\bibitem{mehdi2020exploiting}
M.~Ganji, X.~Zou, and H.~Jafarkhani, ``Exploiting time asynchrony in multi-user
  transmit beamforming,'' \emph{{IEEE} Trans. Wireless Commun.}, vol.~19,
  no.~5, pp. 3156--3169, May 2020.

\bibitem{ding2016design}
Z.~Ding and H.~V. Poor, ``Design of {massive-MIMO-NOMA} with limited
  feedback,'' \emph{{IEEE} Signal Process. Lett.}, vol.~23, no.~5, pp.
  629--633, May 2016.

\bibitem{xu2016outage}
P.~Xu, Y.~Yuan, Z.~Ding, X.~Dai, and R.~Schober, ``On the outage performance of
  non-orthogonal multiple access with 1-bit feedback,'' \emph{{IEEE} Trans.
  Wireless Commun.}, vol.~15, no.~10, pp. 6716--6730, Oct. 2016.

\bibitem{liu2017downlink}
X.~Liu and H.~Jafarkhani, ``Downlink non-orthogonal multiple access with
  limited feedback,'' \emph{{IEEE} Trans. Wireless Commun.}, vol.~16, pp.
  6151--6164, Sep. 2017.

\bibitem{karamad2012quantization}
E.~Karamad, B.~Khoshnevis, and R.~S. Adve, ``Quantization and bit allocation
  for channel state feedback in relay-assisted wireless networks,''
  \emph{{IEEE} Trans. Signal Process.}, vol.~61, no.~2, pp. 327--339, Jan.
  2013.

\end{thebibliography}
}
\end{document}